\documentclass[ aps, showpacs, showkeys, nofootinbib, floatfix, superscriptaddress]{revtex4}

\usepackage{amsfonts}

\usepackage{amssymb}
\usepackage{amsmath}
\usepackage{graphicx}

\begin{document}

\title{The generalized  Brans-Dicke  theory and its cosmology}

 \author{Jianbo Lu}
 \email{lvjianbo819@163.com}
 \affiliation{Department of Physics, Liaoning Normal University, Dalian 116029, P. R. China}
  \author{Yabo Wu}
 \affiliation{Department of Physics, Liaoning Normal University, Dalian 116029, P. R. China}
  \author{Weiqiang Yang}
 \affiliation{Department of Physics, Liaoning Normal University, Dalian 116029, P. R. China}
  \author{Molin Liu}
 \affiliation{College of Physics and Electronic Engineering, Xinyang Normal University, Xinyang, 464000, P. R. China}
\author{Xin Zhao}
 \affiliation{Department of Physics, Liaoning Normal University, Dalian 116029, P. R. China}

\begin{abstract}
 A generalized Brans-Dicke  (GBD) theory is studied in this paper. The GBD theory is obtained by generalizing  the Ricci scalar $R$  to an arbitrary function $f(R)$ in the original Brans-Dicke (BD) action. The interesting property has been found in the GBD theory, for example it can naturally solve the problem of $\gamma$ value emerging in $f(R)$ modified gravity (i.e. the inconsistent problem between the observational $\gamma$ value and the theoretical $\gamma$ value), without introducing the so-called chameleon mechanism.  In this paper, we derive  the cosmological equations and study the cosmology  in the GBD theory. The cosmological  solutions  show that  the GBD model can pass through the test of the observations, such as the observational Hubble data.  Comparing with other theories, it can be found that the GBD theory have some other interesting properties or solve some problems existing in other theories. (1)  It is well known that the $f(R)$ theory are equivalent to the BD theory with a potential (abbreviated as BDV) for taking a specific value of the BD parameter $\omega=0$, where the specific choice: $\omega=0$ for the BD parameter is quite exceptional, and it is hard to understand the corresponding absence of the kinetic term for the field. However, for the GBD theory, it is similar to the double scalar-fields model, and both fields in the GBD own the non-disappeared dynamical effect. (2)  One knows that in the double scalar-fields quintom model, it is required to include both the canonical quintessence field and the non-canonical phantom field  in order to make the state parameter to cross over $w=-1$, while several fundamental problems are associated with phantom field, such as the problem of negative kinetic term and the fine-tuning problem, etc. While, in the GBD model, the state parameter of geometrical dark energy can cross over the phantom boundary $w=-1$ as achieved in the quintom model,  without bearing the problems existing in the quintom model.  (3) The GBD theory tends to investigate the physics from the viewpoint of geometry, while the BDV or the two scalar-fields quintom model tends to solve physical problems from the viewpoint of matter. It is possible that several special characteristics of scalar fields could be revealed through studies of geometrical gravity in the GBD.  As an example, we investigate the potential $V (\phi)$ of the BD scalar field, and an effective form of $V (\phi)$ could be given by studying on the GBD theory. And, it seems that a viable condition for the BD theory could be found, i.e. the BD parameter should be $\omega>0$ for $f>0$, if we assume that the effective form of the BD potential can be approximately written as a popular square function of $\phi$.

\end{abstract}

\pacs{98.80.-k}

\keywords{ Modified gravity; Brans-Dicke theory; Cosmological solution; Effective potential of field.}

\maketitle

\section{$\text{Introduction}$}
 There are several observational and theoretical  motivations to investigate the  modified or alternative theories of general relativity.  Studies on the modified gravity theories of GR have been always the hot area. Several modified gravity theories have been widely studied \cite{mg1,mg2,mg3,mg4,mg5,mg6}, especially two simple  modifications to GR: the $f(R)$ theory \cite{fr-review1,fr-review2} and the  Brans-Dicke (BD)  theory  \cite{original-BD}.

 Recent observations in Refs. \cite{VG-MNRAS-2004-dwarf,VG-PRD-2004-white,VG-APJ,VG-PRD-2002-SN,VG-PRL-1996-neutron} indicate that the Newton gravitational constant $G$ maybe depends on time. Brans-Dicke (BD) theory  is a popular one to describe the time-variable $G(t)$ gravity. As a  simple theory in the scalar-tensor theories \cite{scalar-tensor}, BD theory is apparently compatible with Mach's principle \cite{mach-bd}, and in which a scalar field $\phi$ can be introduced naturally by considering $\phi(t)\propto 1/G(t)$.  But in the original BD theory \cite{original-BD}, it is hard to interpret the cosmic acceleration indicated by the observations \cite{acceleration-98SN,acceleration-99SN,acceleration-WMAP}. In order to obtain an accelerating universe, one usually modified  this theory at three aspects: (1) introducing the invisible component----dark energy in universe \cite{BD-DE}, (2) assuming the coupling constant $\omega$ to be variable with respect to time \cite{BD-omegat1,BD-omegat2}, (3) adding a potential term to the original BD theory (abbreviate as BDV) \cite{BD-potential}.   The applications of these extended BD theories have been investigated widely, such as at the aspects of cosmology \cite{GBD-cosmic1,GBD-cosmic2,GBD-cosmic3}, weak-field approximation \cite{GBD-weak}, observational constraints \cite{GBD-constraint1,GBD-constraint2}, and so on   \cite{BD-widely1,BD-widely2,BD-widely3}.

 In this paper, we investigate other way to explain the cosmic acceleration in the framework of the BD theory, i.e. we generalize the Ricci scalar $R$ to be an arbitrary function $f(R)$ in the original BD action (abbreviate as GBD), which is different from the studies on equivalence between the BD theory and the modified $f(R)$ theory \cite{mg6}.  The interesting property has been found in the GBD model. For example, by using the method of the weak-field approximation Ref. \cite{GBD-L} shows that the GBD theory can naturally solve the problem of $\gamma$ value emerging in $f(R)$ modified gravity (i.e. the inconsistent problem between the observational $\gamma$ value and the theoretical $\gamma$ value), without introducing the so-called chameleon mechanism. The chameleon mechanism is introduced to solve the  problem of $\gamma$ value  in the $f(R)$ modified gravity. Here $\gamma$  is the parametrized post-Newtonian (PPN) parameter.

 The GBD cosmology is studied in this paper, and the structure of our paper is as follows.  In section II, we briefly introduce the GBD theory, and derive to gain  the field equations and the cosmological equations in the GBD theory.  In section III, we give the cosmological solutions of the GBD model.  It is shown that the GBD model can pass through the test of the observation, such as the observational Hubble data.  In section IV and V, we investigate the  properties of the geometrical dark energy and the effective potential of the BD scalar field in the GBD theory. By comparing with the preceding  studies (such as the studies on the $f(R)$, the BDV, and the quintom models), some new ingredients and significant progresses of this work could be shown as follows.  (1) In the GBD theory one can take an arbitrary value of $\omega$ and the kinetic-energy term of scalar field in the action is non-disappeared, which is obviously different from the $f(R)$ theory. The $f(R)$ gravity theory becomes equivalent to the BDV theory for a specific value of  $\omega=0$ under a transformation, where the kinetic-energy term for the scalar field is absent. (2) The GBD theory tends to investigate the physics from the viewpoint of geometry, while the BDV  tends to solve physical problems from the viewpoint of matter. Several special characteristics of scalar fields could be revealed through studies of  geometrical gravity in the GBD, such as we can investigate to given an effective form of potential of the BD scalar field. (3) Comparing with the two scalar-fields quintom model, the effective state parameter of geometrical dark energy in the GBD model can cross over the phantom boundary: $w=-1$  without bearing the problems existing in the non-canonical phantom field. But,  the phantom field is introduced in the double scalar-fields quintom model in order to cross over $w=-1$, where the puzzling  problems  are emergent, such as the  negative kinetic term and the fine-tuning problem. Section VI is the conclusion.

\section{$\text{Field equations and  cosmological equations in the GBD theory}$}

In framework of the time-variable gravitational constant, we study a generalized Brans-Dicke theory by using a function $f(R)$ to replace the Ricci scalar $R$  in the original BD action. The action of system is written as
\begin{equation}
S=S_g(g_{\mu \nu },\phi )+S_m(g_{\mu \nu },\psi )=\frac{1}{2}\int d^4x{\cal L}_{T},\label{action}
\end{equation}
with the total Lagrange quantity ${\cal L}_T$
\begin{equation}
{\cal L}_T ={\cal L}_g+{\cal L}_\phi+{\cal L}_m=\sqrt{-g}[\phi f(R)- \frac{\omega}{2\phi}\partial _\mu \phi \partial ^\mu \phi+\frac{16\pi }{c^4}L_m].\label{lagrange}
\end{equation}
Obviously, the system contains three dynamical variable: the gravitational field $g_{\mu \nu}$, the matter field $\psi$ and the BD scalar field $\phi$. $\omega$ is the couple constant. According to Eq. (\ref{lagrange}), it is easy to see that the GBD theory can be considered as a special case of the more general  $f(R,\phi)$ theory \cite{fR-phi,fR-phi1,fR-phi2}. It is well known that the so-called $f(\phi)R$ theory \cite{fphiR,fphiR1},  as a special case of the $f(R,\phi)$ theory, has been widely studied \cite{fphiR2,fphiR3,fphiR4}. Given that $f(R,\phi)$ is a more complex theory and the more simple theory is usually more favored by the researcher in physics, here we investigate the GBD model induced  by  the directly observational motivation of the accelerating universe and some other motivations exhibited in the introduction. Concretely, we discuss some interesting cosmological contents in the GBD model, such as the comparison with observation, the properties of effective state parameter for the geometrical dark energy,  the effective potential of the BD scalar field, etc.

Taking $c=1$ and varying the action with respect to metric
$\frac{\delta S}{\delta g^{\mu \nu }}=\frac{\delta S_g(g_{\mu \nu },\phi )}{\delta g^{\mu \nu }}+\frac{\delta S_m(g_{\mu \nu },\psi )}{\delta g^{\mu \nu}}=0$, one can get the gravitational field equation
\begin{eqnarray}
\phi \left[ f_{R}R_{\mu \nu }-\frac{1}{2}f(R)g_{\mu \nu }\right]- (\nabla_\mu\nabla_\nu -g_{\mu\nu}\Box)(\phi f_{R})+ \frac{1}{2}\frac{\omega}{\phi}g_{\mu\nu}\partial_\sigma\phi\partial^\sigma\phi
-\frac{\omega}{\phi}\partial_\mu\phi\partial_\nu\phi = 8\pi T_{\mu \nu },\label{gravitational-eq}
\end{eqnarray}
where $f_{R}\equiv \partial f/\partial R$, $\nabla _\mu $ is the covariant derivative associated with the Levi-Civita connection of the metric,  $\Box \equiv \nabla ^\mu \nabla _\mu $, and $T_{\mu \nu }=\frac{-2}{\sqrt{-g}}\frac{\delta S_m}{\delta g^{\mu \nu }}$ is the energy momentum tensor of the matter. Varying the action (\ref{action}) with respect to the scalar field $\phi$ and the matter field  $\psi$   give respectively
\begin{equation}
f(R)+2\omega\frac{\Box \phi}{\phi} -\frac{\omega}{\phi^{2}}\partial _\mu \phi \partial ^\mu \phi=0,\label{scalar-eq}
\end{equation}
 \begin{equation}
 \frac{\delta S}{\delta \psi}=\frac{\delta S_g(g_{\mu \nu },\phi )}{\delta \psi}+\frac{\delta S_m(g_{\mu \nu },\psi )}{\delta \psi}=0.
 \end{equation}
The trace of Eq. (\ref{gravitational-eq}) is
\begin{eqnarray}
f_{R}R-2f(R)+\frac{3\Box (\phi f_{R})}{\phi}+\frac{\omega}{\phi^{2}}\partial_\mu\phi\partial^\mu\phi
= \frac{8\pi T}{\phi}.\label{trace}
\end{eqnarray}
From Eqs. ({\ref{scalar-eq}) and (\ref{trace}), one can see that the curvature of the spacetime could be caused by the motion of $\phi$.  And from Eq. (\ref{gravitational-eq}), it is shown  that the BD scalar field does not exert any direct influence on matter,  while it couples with another scalar field $f_{R}$. Furthermore, the standard $f(R)$ modified gravity is recovered for $\phi$=constant, while above equations reduce to the Einstein's general relativity (GR) for both BD scalar field $\phi$=constant and $f(R)=R$. Combining Eqs. (\ref{scalar-eq}) and (\ref{trace}), we get
 \begin{eqnarray}
\Box \phi-\frac{\partial_{\mu}\phi\partial ^{\mu}\phi}{4\phi}=\frac{1}{4\omega}[8\pi T -\phi R f_{R}-3\Box (\phi f_{R})].\label{dynamical-phi}
\end{eqnarray}
One can read from Eq. (\ref{dynamical-phi}) that, for $\omega\rightarrow \infty$ the constant-$G$ theory can  be recovered, which is same to the result in the standard BD theory.

In the flat Friedmann-Lemaitre-Robertson-Walker (FLRW) metric
\begin{equation}
ds^2=-dt^2+a^2\left( t\right) d\vec{x}^2,
\end{equation}
using Eqs. (\ref{gravitational-eq}) and (\ref{scalar-eq}), we can derive the  evolutional equations of the background universe in the GBD theory,
\begin{equation}
3f_{R}H^2=\frac{8\pi\rho_{m}}{\phi}+\frac{f_{R}R-f(R)}{2}-3H\dot{f_{R}}+\frac{1}{2}\omega\left(\frac{\dot{\phi}}{\phi}\right)^{2}
-3Hf_{R}\frac{\dot{\phi}}{\phi},\label{tt-component}
\end{equation}
\begin{equation}
-2f_{R}\dot{H}=\frac{8\pi}{\phi}(\rho_{m}+p_{m})+\ddot{f_{R}}-H\dot{f_{R}}+\omega\left(\frac{\dot{\phi}}{\phi}\right)^{2}
-Hf_{R}\frac{\dot{\phi}}{\phi}+f_{R}\frac{\ddot{\phi}}{\phi}+2\frac{\dot{\phi}}{\phi}\dot{f_{R}},\label{ss-component}
\end{equation}
\begin{equation}
f(R)-\omega\left(\frac{\dot{\phi}}{\phi}\right)^{2}+2\omega\frac{\stackrel{\cdot \cdot }{\phi } }\phi +6\omega H \frac{\dot{\phi}}{\phi}=0.\label{phi-motion-fRBD}
\end{equation}
 Here $a$ is the cosmic scale factor, $H$ is the Hubble parameter, $R=6\left( 2H^2+\stackrel{\cdot }{H}\right)$, and "dot" denotes the derivative with respect to cosmic time $t$. For case of $\phi$=constant ($\dot{\phi}=0$ and $\ddot{\phi}=0$) in Eqs.(\ref{tt-component}-\ref{phi-motion-fRBD}), they are reduced to the $f(R)$ theory, while for case of $f(R)=R$ they are reduced to the original Brans-Dicke theory.

\section{$\text{ Cosmological  solutions in the GBD theory}$}

For solving the cosmological equations (\ref{tt-component}-\ref{phi-motion-fRBD}), we define the dimensionless variables:
\begin{equation}
y_{H}=H^{2}/m^{2}-a^{-3},\label{dimensionless-H}
\end{equation}
\begin{equation}
y_{R}=R/m^{2}-3a^{-3},\label{dimensionless-R}
\end{equation}
\begin{equation}
y_{\phi}=\phi/\phi_{0},\label{dimensionless-phi}
\end{equation}
\begin{equation}
y_{\phi}^{'}=\phi^{'}/\phi_{0},\label{dimensionless-phip}
\end{equation}
Thus using Eqs. (\ref{tt-component}) and (\ref{phi-motion-fRBD}), we get the differential equations for $\{y_{H},y_{R},y_{\phi},y_{\phi}^{'}\}$ as follows
\begin{equation}
y_{H}^{'}=\frac{1}{3}y_{R}-4y_{H},\label{yH-p}
\end{equation}
\begin{equation}
y_{R}^{'}=\frac{-[(y_{H}+a^{-3})f_{R}-\frac{f_{R}}{6}(y_{R}+3a^{-3})+\frac{f}{6m^{2}}-
\frac{\omega}{6}(\frac{y_{\phi}^{'}}{y_{\phi}})^{2}(y_{H}+a^{-3})+f_{R}\frac{y_{\phi}^{'}}{y_{\phi}}(y_{H}+a^{-3})-\frac{a^{-3}}{\phi}]}{(y_{H}+a^{-3})m^{2}f_{RR}}
+9a^{-3},\label{yR-p}
\end{equation}
\begin{equation}
y_{\phi}^{'}=\phi^{'}/\phi_{0},\label{yphi-p}
\end{equation}
\begin{equation}
y_{\phi}^{''}=\frac{\phi}{2\omega(y_{H}+a^{-3})}[\frac{-f}{m^{2}}+\omega(\frac{y_{\phi}^{'}}{y_{\phi}})^{2}(y_{H}+a^{-3})
-\omega\frac{y_{\phi}^{'}}{y_{\phi}}(\frac{1}{3}y_{R}-4y_{H}-3a^{-3})-6\omega\frac{y_{\phi}^{'}}{y_{\phi}}(y_{H}+a^{-3})].\label{yphi-pp}
\end{equation}
Here the subscript "0" denotes the current value of parameters, the superscript $'$ denotes the derivative with respect to $\ln a$, the parameter $m$ is defined as  $m^{2}=(8315 Mpc)^{-2}(\frac{\Omega_{0m}h^{2}}{0.13})$ and $\Omega_{0m} $ is the current dimensionless energy density of the matter.  To solve above  differential equations, the initial conditions ($a_{0}=1$) are expressed respectively as
\begin{equation}
y_{H}|_{a=1}=H_{0}^{2}/m^{2}-1,\label{initial-H}
\end{equation}
\begin{equation}
y_{R}|_{a=1}=6H_{0}^{2}(1-q_{0})/m^{2}-3,\label{initial-q}
\end{equation}
\begin{equation}
y_{\phi}|_{a=1}=1,\label{initial-phi}
\end{equation}
\begin{equation}
y_{\phi}^{'}|_{a=1}=0.01.\label{initial-phip}
\end{equation}
Here $q=-\frac{\ddot{a}}{aH^{2}}$ is the deceleration parameter, and its current value $q_{0}$ can be given by the cosmic observations. The value of the initial condition $y_{\phi}^{'}|_{a=1}$ can be indicated by the following observations. For example, the limits on the variation of $G$ can be exhibited by: $|\frac{\dot{G}}{G}|=|\frac{\dot{\phi}}{\phi}|\leq 4.1\times 10^{-10} y^{-1}$ from Pulsating white dwarf G117-B15A \cite{VG-MNRAS-2004-dwarf}, $-4\times 10^{-11} y^{-1} \leq \frac{\dot{\phi}}{\phi} \leq 2.5\times 10^{-10} y^{-1}$ from Nonradial pulsations of white dwarfs \cite{VG-PRD-2004-white}, $|\frac{\dot{\phi}}{\phi}|\leq 2.3\times 10^{-11} y^{-1}$ from Millisecond pulsar PSR J0437-4715 \cite{VG-APJ}, $|\frac{\dot{\phi}}{\phi}|\leq  10^{-11} y^{-1}$ from Type-Ia supernovae \cite{VG-PRD-2002-SN}, $\frac{\dot{\phi}}{\phi}= (0.6\pm 4.2)\times 10^{-12} y^{-1}$ from Neutron star masses \cite{VG-PRL-1996-neutron}, $|\frac{\dot{\phi}}{\phi}|\leq 1.6\times 10^{-12} y^{-1}$ from Helioseismology \cite{vg-constraint6}, and $\frac{\dot{\phi}}{\phi}= (4\pm 9)\times 10^{-13} y^{-1}$ from Lunar laser ranging experiment \cite{vg-constraint7}, etc. Taking a stringent bound $|\frac{\dot{\phi}}{\phi}|\leq  10^{-12} y^{-1}$ and  considering the current value of the dimensionless Hubble constant $h = 0.673\pm0.010$ from the Planck 2015 results \cite{hubble-value},   we can calculate to limit $|y_{\phi}^{'}({a=1})|\leq 0.015$ by using the center value  $H_{0}= 67.3km s^{-1} Mpc^{-1} = 6.87 \times  10^{-11} y^{-1}$. Here we take $y_{\phi}^{'}({a=1})= 0.01$ as an initial condition in Eq. (\ref{initial-phip}). For comparison, the cases of other initial values of  $y_{\phi}^{'}({a=1})$ (less than 0.01) are also discussed.

 To find a cosmological solution of the GBD theory, we need to take a concrete form of  $f(R)$ function at prior.   As an example, we consider an interesting model called ¡°exponential gravity¡±
\begin{equation}
f(R)=R-\beta R_{s}(1-e^{-R/R_{s}}),\label{fR-e}
\end{equation}
 which is proposed by Refs.\cite{viable-fr-e1,viable-fr-e2,viable-fr-e3}. Here $\beta$  and $R_{s}$ are two constants with $\beta R_{s}\simeq 12 H_{0}^{2}\Omega_{0m}$ \cite{,viable-fr-e3}. This model  has an important feature that it has only one more parameter than the $\Lambda$CDM model. The first and the second derivatives of Eq. (\ref{fR-e}) with respect to $R$ are
 \begin{equation}
f_{R}=1-\beta e^{-R/R_{s}},\label{fR-e-D1}
\end{equation}
\begin{equation}
f_{RR}=\frac{\beta}{R_{s}}e^{-R/R_{s}}.\label{fR-e-D2}
\end{equation}
\begin{figure}[ht]
  \includegraphics[width=7.5cm]{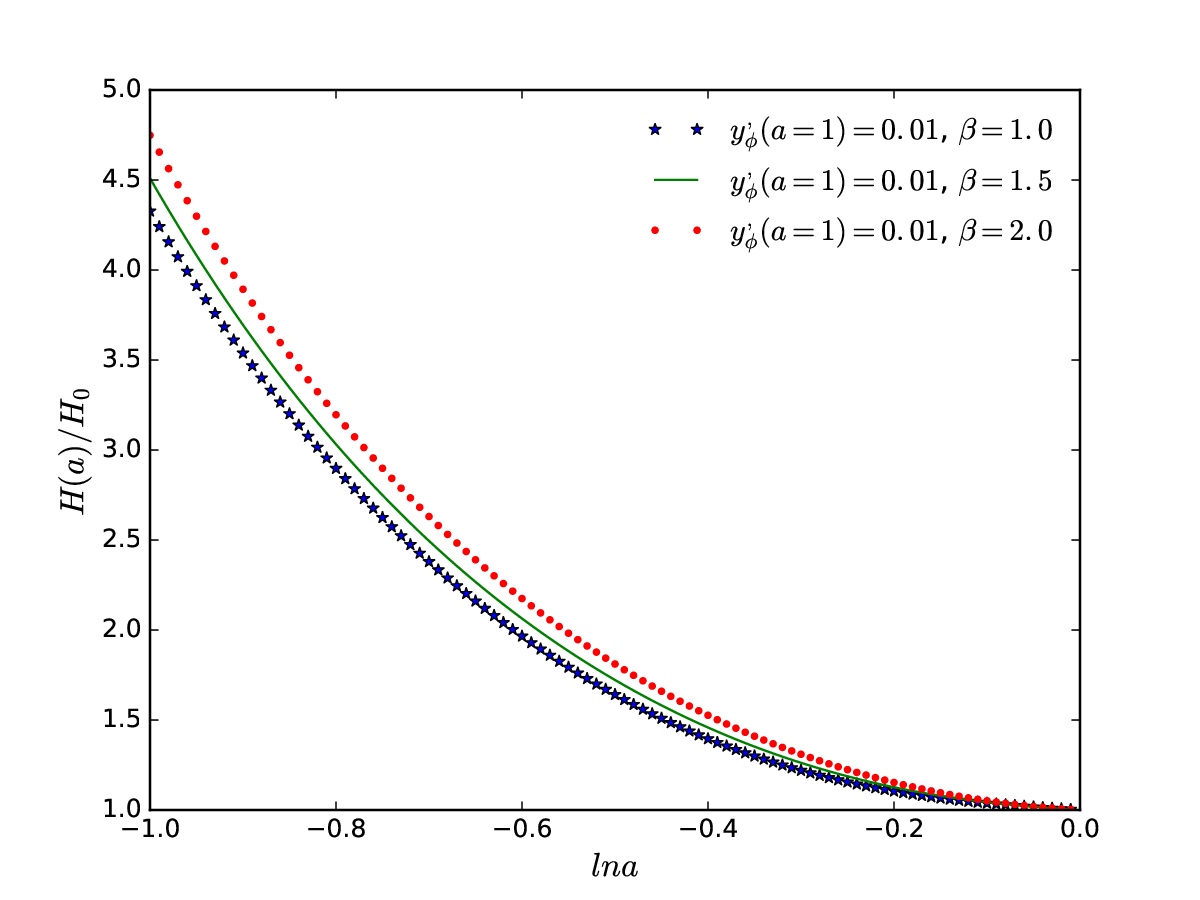}
  \includegraphics[width=5.5cm]{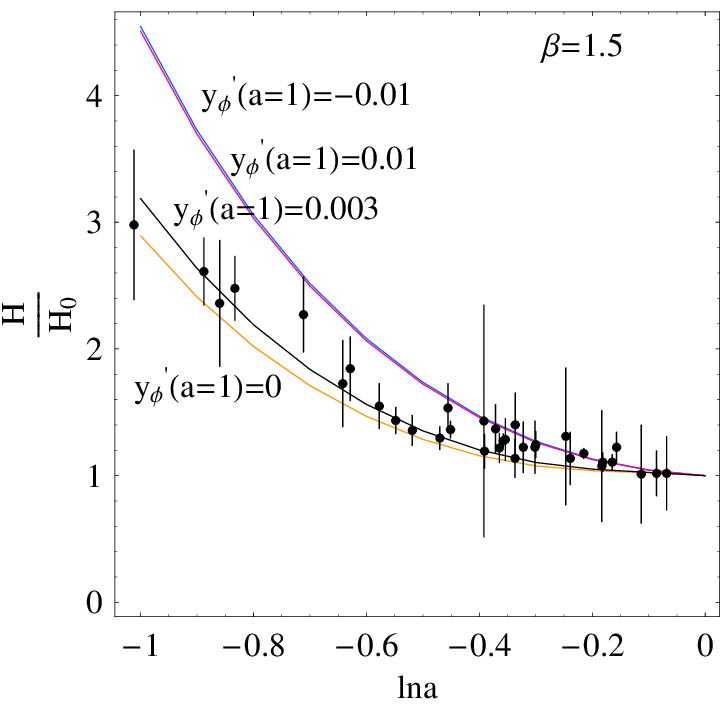}\\
  \caption{The numerical solutions of $H(a)/H_{0}$ in the GBD model with the different model parameter $\beta$ or the  different  initial condition  $y_{\phi}^{'}(a_{0})$. }\label{figure-E}
\end{figure}
\begin{figure}[ht]
  \includegraphics[width=7.5cm]{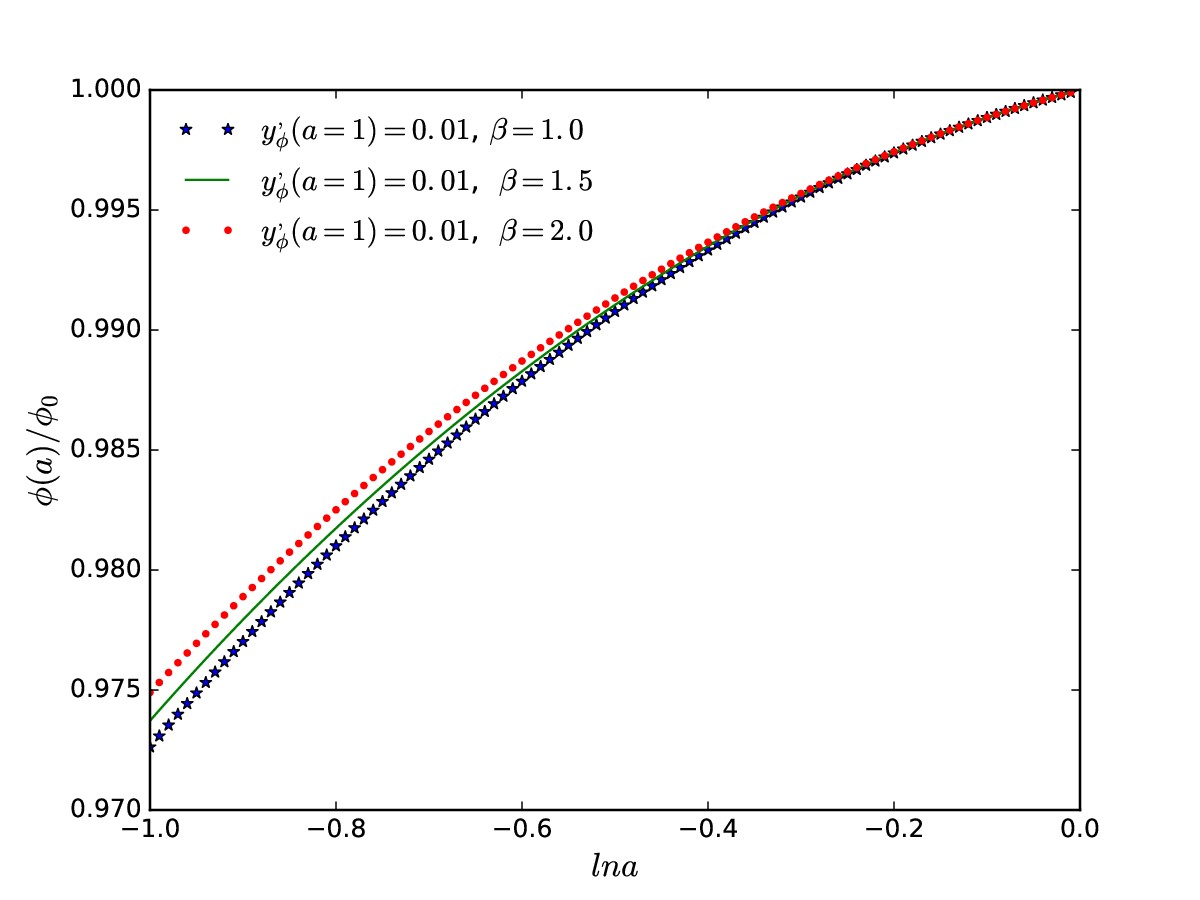}
  \includegraphics[width=7.5cm]{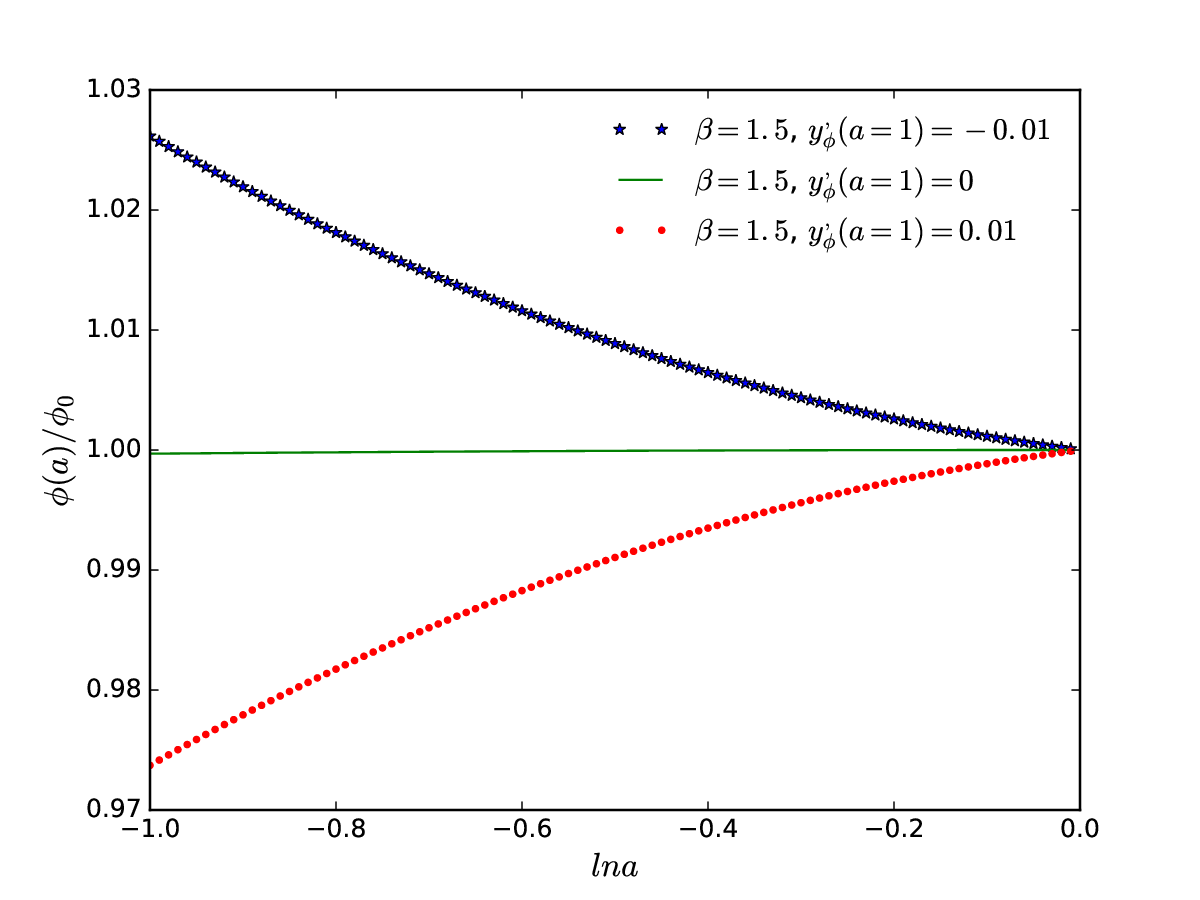}\\
  \caption{The numerical solutions of $\phi(a)/\phi_{0}$  in the GBD model  with the different model parameter $\beta$ or the  different initial condition  $y_{\phi}^{'}(a_{0})$. }\label{figure-phi}
\end{figure}
Thus using the system of the ordinary differential equations (\ref{yH-p})-(\ref{yphi-pp}) and the initial conditions (\ref{initial-H})-(\ref{initial-phip}), we can numerically exhibit the solutions: $H(a)$ and $\phi(a)$ in the GBD theory, which are illustrated in Fig.\ref{figure-E} and Fig.\ref{figure-phi}.

In Fig.\ref{figure-E} (left), we show the dependence of $H(a)$ on the parameter $\beta$.  Fig.\ref{figure-E} (right) illustrates  the evolution of $H(a)$ with respect to $\ln a$ with taking the different values of $y_{\phi}^{'}(a=1)$. In the following, we use $a_{0}$ to denotes the current value: $a=1$. We can see  that the evolutions of $H(a)$ almost have the same trajectory for two cases: $y_{\phi}^{'}(a_{0})=0.01$ and $y_{\phi}^{'}(a_{0})=-0.01$, while the evolutions of $H(a)$ are obviously different for two cases: $y_{\phi}^{'}(a_{0})=0.01$ and $y_{\phi}^{'}(a_{0})=0$. It seems that the effect to $H(a)$ from the BD field is notable. Using the observational Hubble data listed in table \ref{table-ohd}, we display these observational $H(z)$ value  in Fig.\ref{figure-E} (right). Here $z=(1-a)/a$ is the cosmic redshift. It is shown from Fig.\ref{figure-E} (right) that the most observational $H(z)$ data are located in the region between the case of $y_{\phi}^{'}(a_{0})=0$ and the case of $y_{\phi}^{'}(a_{0})=\pm 0.01$. It seems that  the GBD model could pass through the test of the observation, such as the observational Hubble data, since the evolution of  $H(a)$ with  $y_{\phi}^{'}(a_{0})=0.003$   is well consistent with those observational data. From Fig.\ref{figure-phi} (right), one can see that evolutional tendency of BD scalar field depends on the initial value of  $y_{\phi}^{'}(a_{0})$.

\begin{table}[!htbp]
 \vspace*{-12pt}
 \begin{center}
 \begin{tabular}{ |c|| c| c| c |c| c| c |c| c| c | } \hline\hline
 z     & 0.0708           & 0.09               &  0.12             & 0.17             & 0.179           & 0.199          &0.20         &0.24& 0.27\\\hline
 H(z)  &$69.0 \pm 19.68$  & $69.0 \pm 12.0$    &  $68.6 \pm 26.2$  &  $83.0 \pm 8.0$  &  $75.0 \pm 4.0$ & $75.0 \pm 5.0$ &$72.9 \pm 29.6$ & $79.69 \pm 2.65$
 & $77.0 \pm 14.0$ \\\hline
Ref.  & \cite{ohd-13}  & \cite{ohd-14}   &\cite{ohd-13}  &\cite{ohd-15} &\cite{ohd-16} &\cite{ohd-16} &\cite{ohd-13}& \cite{ohd-17}&\cite{ohd-15} \\\hline
z    & 0.28             & 0.35           & 0.352              &  0.3802             & 0.4             & 0.4004          & 0.4247       &0.43      &0.44\\\hline
 H(z) & $88.8 \pm 36.6$  &$84.4 \pm 7.0$  & $83.0 \pm 14.0$    &  $83.0 \pm 13.5$  &  $95.0 \pm 17.0$  &  $77.0 \pm 10.2$ & $87.1 \pm 11.2$ &$86.45 \pm 3.68$ & $82.6 \pm 7.8$ \\\hline
Ref.  & \cite{ohd-13}   &\cite{ohd-18}  &\cite{ohd-16}  &\cite{ohd-19} &\cite{ohd-15} &\cite{ohd-19} &\cite{ohd-19}&\cite{ohd-17}&\cite{ohd-20} \\\hline
z    & 0.4497           & 0.4783         & 0.48               & 0.57              &  0.593          & 0.6          & 0.68      & 0.73       &0.781    \\\hline
 H(z)  & $92.8 \pm 12.9$ & $80.9 \pm 9.0$ &$97.0 \pm 62.0$  & $92.4 \pm 4.5$ &  $104.0 \pm 13.0$  &  $87.9 \pm 6.1$  &  $92.0 \pm 8.0$ & $97.3 \pm 7.0$ &$105.0 \pm 12.0$ \\\hline
Ref.  & \cite{ohd-19}  &\cite{ohd-19} &\cite{ohd-21} &\cite{ohd-22} &\cite{ohd-16} &\cite{ohd-20} &\cite{ohd-16}&\cite{ohd-20}&\cite{ohd-16} \\\hline
 z      &0.875             & 0.9               & 1.037           & 1.3              &  1.363            & 1.43        & 1.75  && \\\hline
  H(z)  &  $125.0 \pm 17.0$   & $117.0 \pm 23.0$ &$154.0 \pm 20.0$  & $168.0 \pm 17.0$    &  $160.0 \pm 33.6$  &  $177.0 \pm 18.0$ & $202.0 \pm 40.0$ &&\\\hline
Ref.  & \cite{ohd-16}   &\cite{ohd-15}  &\cite{ohd-16} &\cite{ohd-15} &\cite{ohd-23} &\cite{ohd-15}&\cite{ohd-15} && \\\hline
 \hline
 \end{tabular}
 \end{center}
 \caption{The values of Hubble parameter given by observations. }\label{table-ohd}
 \end{table}

\section{$\text{Effective state parameter of geometrical dark energy in the GBD}$}

Probing properties of the dark energy is important, and it has been studied in the standard cosmology or the several modified gravity theories \cite{DE,DE1,DE2,DE3,DE4,DE5,DE6,DE7,DE8,DE9,DE10,DE11,multi-field,multi-field1,multi-field2,DE-lu1,DE-lu2}. Next we investigate the properties of geometrical dark energy in this GBD theory, and analyze the effects of the BD scalar field. Rewriting the Eq.(\ref{gravitational-eq}) as follows
\begin{eqnarray}
G_{\mu\nu}&=&R_{\mu\nu}-\frac{1}{2}Rg_{\mu\nu}=\frac{8\pi T_{\mu \nu }^{eff}}{\phi_{0}},\label{gravitational-eq1}
\end{eqnarray}
with
\begin{eqnarray}
&&T_{\mu \nu }^{eff}=\frac{\phi_{0}}{\phi}\frac{T_{\mu\nu}}{f_{R}} \nonumber \\
&&+\frac{\phi_{0}}{8\pi f_{R}}[(\nabla_\mu \nabla _\nu -g_{\mu \nu }\Box )f_{R} +\frac{1}{2}f(R)g_{\mu \nu }-\frac{1}{2}f_{R}Rg_{\mu\nu}+ \frac{f_{R}}{\phi}(\nabla_\mu\nabla_\nu -g_{\mu\nu}\Box)\phi \nonumber \\
&&- \frac{1}{2}\frac{\omega}{\phi^{2}}g_{\mu\nu}\partial_\sigma\phi\partial^\sigma\phi +\frac{\omega}{\phi^{2}}\partial_\mu\phi\partial_\nu\phi],\label{Tmunu}
\end{eqnarray}
then the effective energy density and the effective pressure are derived as
\begin{eqnarray}
\rho^{eff}=
\frac{\phi_{0}}{\phi}\frac{\rho_{m}}{f_{R}}+\frac{\phi_{0}}{8\pi f_{R}} \left[-3H\dot{f_{R}}-\frac{1}{2}f(R)+\frac{1}{2}f_{R}R-3Hf_{R}\frac{\dot{\phi}}{\phi}
+\frac{1}{2}\omega(\frac{\dot{\phi}}{\phi})^{2}\right]  \label{rho-eff}
\end{eqnarray}
\begin{eqnarray}
p^{eff}=\frac{\phi_{0}}{\phi}\frac{p_{m}}{f_{R}}+\frac{\phi_{0}}{8\pi f_{R}} \left[\ddot{f_{R}}+2H\dot{f_{R}}+\frac{1}{2}f(R)-\frac{1}{2}f_{R}R+f_{R}\frac{\ddot{\phi}}{\phi}
+2Hf_{R}\frac{\dot{\phi}}{\phi} +\frac{1}{2}\omega(\frac{\dot{\phi}}{\phi})^{2}+2\frac{\dot{\phi}}{\phi}\dot{f_{R}}\right]. \label{p-eff}
\end{eqnarray}
Here $\rho_{m}$ and $p_{m}$ are the energy density and the pressure of matter, respectively. According to Eqs. (\ref{gravitational-eq1}) and (\ref{Tmunu}), we can define the effective Newton gravitational constant $G_{eff}=\frac{1}{\phi f_{R}}$. To keep the attractive property of gravity, we get an constraint: $\phi f_{R}>0$. If we assume $\phi>0$, then $f_{R}>0$. The effective  state parameter  for geometrical dark energy has a form
\begin{eqnarray}
w^{eff}_{g}=\frac{p^{eff}-p_{m}}{\rho^{eff}_{g}-\rho_{m}}=\frac{8\pi \phi_{0} p_{m}-8\pi \phi p_{m}f_{R}+\phi_{0}\phi\left[\ddot{f_{R}}+2H\dot{f_{R}}+\frac{1}{2}f(R)-\frac{1}{2}f_{R}R+f_{R}\frac{\ddot{\phi}}{\phi}
+2Hf_{R}\frac{\dot{\phi}}{\phi} +\frac{1}{2}\omega(\frac{\dot{\phi}}{\phi})^{2}+2\frac{\dot{\phi}}{\phi}\dot{f_{R}}\right]}{8\pi \phi_{0} \rho_{m}-8\pi \phi \rho_{m}f_{R}+\phi_{0}\phi\left[-3H\dot{f_{R}}-\frac{1}{2}f(R)+\frac{1}{2}f_{R}R-3Hf_{R}\frac{\dot{\phi}}{\phi}
+\frac{1}{2}\omega(\frac{\dot{\phi}}{\phi})^{2}\right]}.\label{wde-eff}
\end{eqnarray}

\begin{figure}[ht]
  \includegraphics[width=7.5cm]{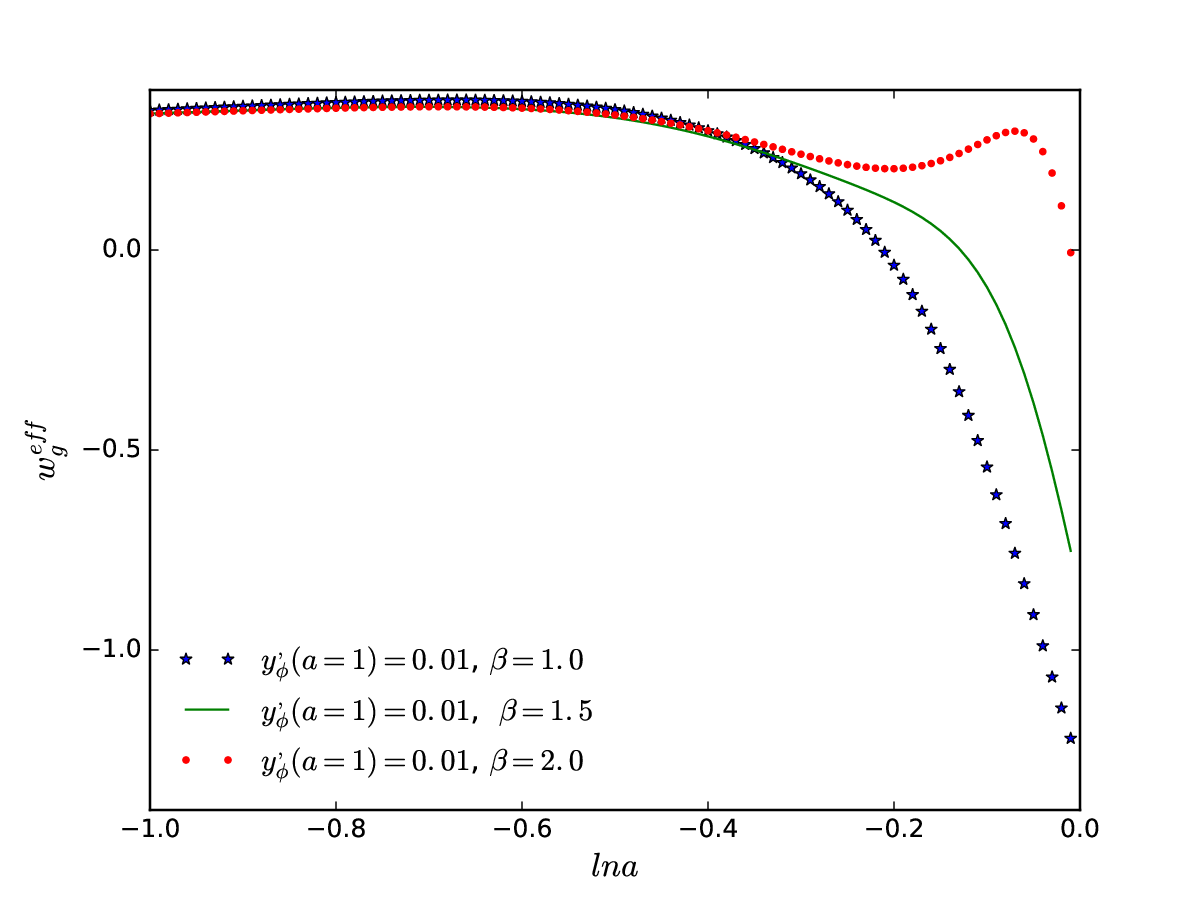}
  \includegraphics[width=7.5cm]{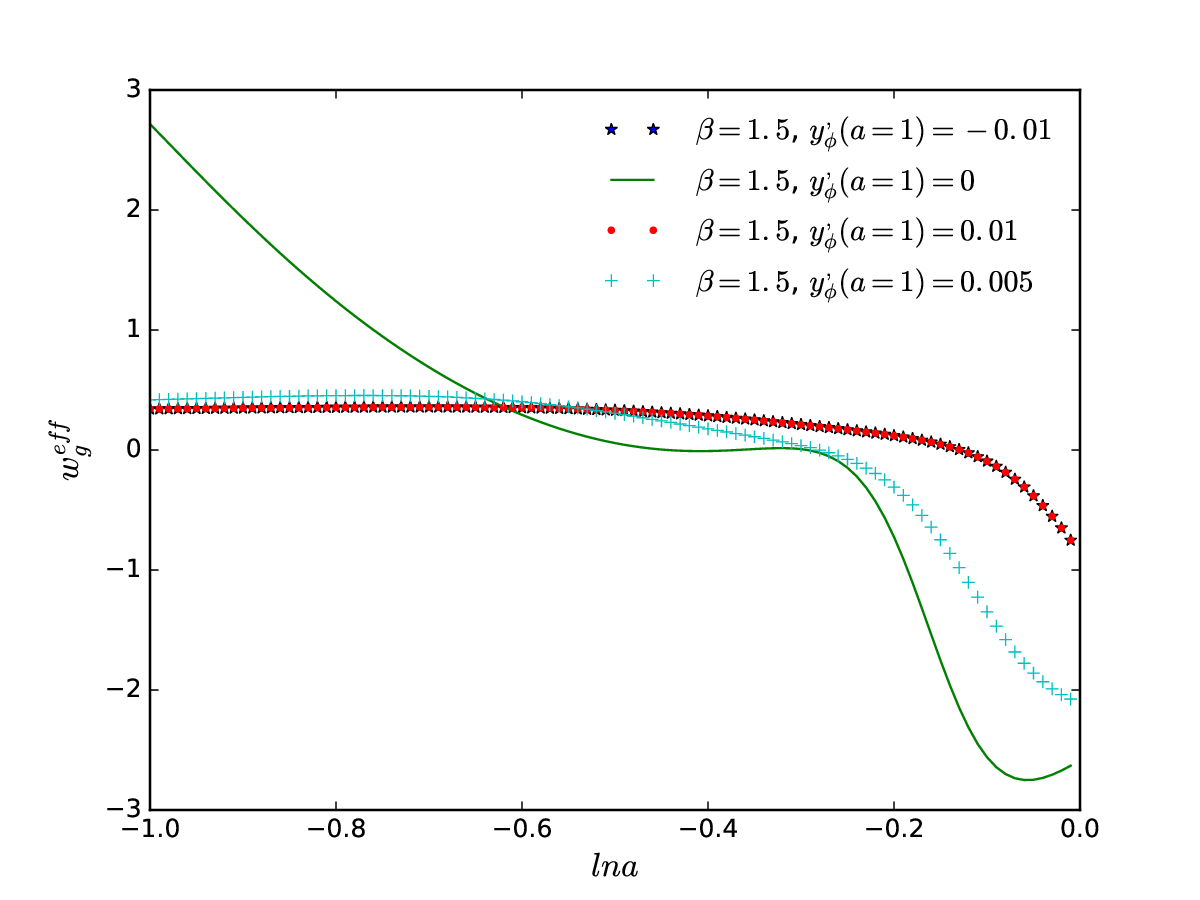}\\
  \caption{The evolutions of $w_{g}^{eff}(a)$ in the GBD model  with the different model parameter $\beta$ or the  different initial condition  $y_{\phi}^{'}(a_{0})$. }\label{figure-weos}
\end{figure}

 Taking the function $f(R)=R+\beta R_{s}(1-e^{-R/R_{s}})$ as an example, we plot the evolution of $w^{eff}_{g} (a)$  in Fig.\ref{figure-weos}  by using the different values of model parameter $\beta$ and the initial values $y_{\phi}^{'}(a_{0})$. In this GBD model, the dependence of $w^{eff}_{g}  (a)$ on the model parameter $\beta$ are illustrated in Fig.\ref{figure-weos} (left).  In the Fig.\ref{figure-weos} (right), one can see that  $w^{eff}_{g} (a)$ almost have the same evolutions for the two cases: $y_{\phi}^{'}(a_{0})=\pm 0.01$, i.e. the trajectories of $w^{eff}_{g} (a)$ are not sensitive to the symbol of initial condition $y_{\phi}^{'}(a_{0})$,  while the effect on  $w^{eff}_{g} (a)$ from the BD scalar field is notable since  the evolution of $w^{eff}_{g} (a)$ with $y_{\phi}^{'}(a_{0})=0$ is obviously different from other three cases: $y_{\phi}^{'}(a_{0})=\pm 0.01$ and $y_{\phi}^{'}(a_{0})=0.005$. Also, one can see that the current value $w^{0eff}_{g}$ with $y_{\phi}^{'}(a_{0})=0$  has the more small value than other cases of  $y_{\phi}^{'}(a_{0})\neq 0$, while  $w^{eff}_{g} (a)$ with $y_{\phi}^{'}(a_{0})=0$ has the more large value than the cases of  $y_{\phi}^{'}(a_{0})\neq 0$ at the high redshift $\ln a=-1$. And the values of $w^{0eff}_{g}$  are located in the range [-2.63,-0.75] for using the different initial conditions: from $y_{\phi}^{'}(a_{0})=0$ to $y_{\phi}^{'}(a_{0})=\pm 0.01$.  The evolutions of $w^{eff}_{g} (a)$ with $y_{\phi}^{'}(a_{0})=\pm 0.01$  in Fig.\ref{figure-weos}  show  that they vary from $w\sim\frac{1}{3}$ (radiation) to $w<0$ (dark energy).

 It is shown in Eqs.(\ref{scalar-eq}) and (\ref{trace}) that the GBD model can be equivalently considered as the two scalar-fields model. It can be found in  Fig.\ref{figure-weos} that  the effective state parameter of geometrical dark energy  in the GBD model can cross over the phantom boundary, as achieved in the double scalar-fields  quintom model  \cite{quintom-feng}. But  we can notice that  it is required to include both the canonical quintessence field and the non-canonical phantom field  in the double scalar-fields  quintom model  \cite{quintom-feng}, in order to make the state parameter to cross over the phantom boundary: $w=-1$, where several fundamental problems are associated with phantom field, such as the problem of negative kinetic term and the fine-tuning problem, etc. It is shown that the GBD model can paly a role of the quintom  without bearing the problems existing in the two-fields quintom model, which is also a motivation to appeal us to study the GBD model.

\section{$\text{ A effective potential of the BD scalar field in the GBD theory}$}

 One knows that the potential of a scalar field usually paly an important role in the early inflation universe and the late accelerating universe.  Determining the forms of the potential function for a scalar field is significative, since the potential display some properties for a scalar field.
 By using Eqs. (\ref{scalar-eq}) and (\ref{trace}), the equations of motion for the Brans-Dicke scalar field $\phi$ and the  other scalar field $\varphi=f_{R}$  are expressed as follows
 \begin{eqnarray}
\Box \phi=\frac{\partial_{\mu}\phi\partial^{\mu}\phi}{2\phi}-\frac{\phi f}{2\omega}=-V_{\phi}(\phi),\label{vphip}
\end{eqnarray}
\begin{eqnarray}
\Box f_{R}=\frac{1}{3}[\frac{8\pi T}{\phi}-f_{R}R+2f-3f_{R}
\frac{\Box \phi}{\phi}-6\dot{f_{R}}\frac{\dot{\phi}}{\phi}-\frac{\omega\partial_{\mu}\phi\partial^{\mu}\phi}{\phi^{2}}]
=- \mathcal{V}_{\varphi} (\varphi)+\frac{8\pi T}{3\phi}.\label{vpship}
\end{eqnarray}
 Obviously, this GBD theory is similar to the two scalar-fields theory. Here $V(\phi)$ and $\mathcal{V}(\varphi)$ denotes the effective potential of fields, and the subscript $\phi$ (or $\varphi$) denotes the derivative with respect to scalar field. The effective forms of $V_{\phi}$ and $\mathcal{V}_{\varphi}$ can be gained by comparing with the standard form of equation of motion for the  canonical scalar field, i.e. the standard  Klein-Gordon equation in the vacuum: $\Box \Phi+V_{\Phi}(\Phi)=0$.

  One knows that the geometrical representation may be more appealing to relativists due to its more apparent geometrical nature, whereas the scalar-field representation seems more appealing to particle physicists. Obviously, the GBD theory tends to investigate the physics from the viewpoint of geometry, while the BDV tends to solve physical problems from the viewpoint of matter. Given that the equivalence between the BDV theory and the $f(R)$ theory, some properties of the BD scalar field could be found. So, it is possible that several special characteristics of scalar fields could be revealed through studies of geometrical gravity in the GBD. Next we investigate the effective form of  potential $V(\phi)$ of the BD scalar field $\phi$. Assuming that the variable $\phi$ is independent of its derivative $\partial_{\mu}\phi$ and the geometrical quantity $f(R)$,   we can gain an effective form of $V(\phi)$ by integrating  Eq.(\ref{vphip}) with respect to $\phi$
\begin{eqnarray}
V(\phi)=\frac{H^{2}\phi^{'2}}{2}\ln\phi+\frac{f}{4\omega}\phi^{2}+C(\dot{\phi},f).\label{V-phi}
\end{eqnarray}
Here  superscript $'$ denotes the derivative with respect to $\ln a$, $C$ is a parameter that is independent of $\phi$. Using Eq.(\ref{V-phi}) and taking $C=0$, we can plot the shapes of BD effective potential in Fig.\ref{figure-vphi}. We can see from Fig.\ref{figure-vphi} that the  trajectories of BD effective potential $V(\phi)$ are not sensitive to the variation of $\beta$ values, while the shapes of $V(\phi)$ much depend on the initial condition $y_{\phi}^{'}(a_{0})$ for the smaller $\ln a$ ($\ln a <-0.4$), and for $\ln a>-0.4$ one has $V(\phi)\sim 0$. Obviously, C$(\dot{\phi},f)$ in Eq. (\ref{V-phi}) is a undetermined freedom, whose uncertainty can be used to modify the trajectories of the BD effective potential.

Furthermore,  an interesting property can be found in the GBD theory by comparing with $f(R)$ theory.   It is well known that, the $f(R)$ theory are equivalent to the BDV theory with taking a specific value of the BD parameter $\omega=0$ \cite{fr-bd2,fr-bd3}. However, the specific choice: $\omega=0$ for the BD parameter is quite exceptional, and it is hard to understand the corresponding absence of the kinetic-energy term for the scalar field.  But in the GBD theory,   one can read  from Eq.(\ref{lagrange}) that the value of $\omega$ is arbitrary and the kinetic-energy term of the scalar field is non-disappeared. In addition, if we assume that the effective form of the BD potential can be  approximately written as a popular square function of $\phi$, i.e. we assume  $V(\phi)\simeq V_{0}+\frac{1}{2}m^{2}\phi^{2}$ \cite{bd-potential1,bd-potential2}, then we have $V_{\phi}\simeq m^{2}\phi$ with $m$ owning the mass dimension. Thus, we need to require $\dot{\phi}^{2}\simeq 0$ (i.e. a slow-rolling field) in Eq. (\ref{vphip}), and then we gain $m^{2}=\frac{f}{2\omega}$. Obviously, if $f>0$, we get an constraint on the BD parameter $\omega>0$. And $\omega$ should own a large value with the requirement of a small value of $m^{2}$.

\begin{figure}[ht]
  \includegraphics[width=7.5cm]{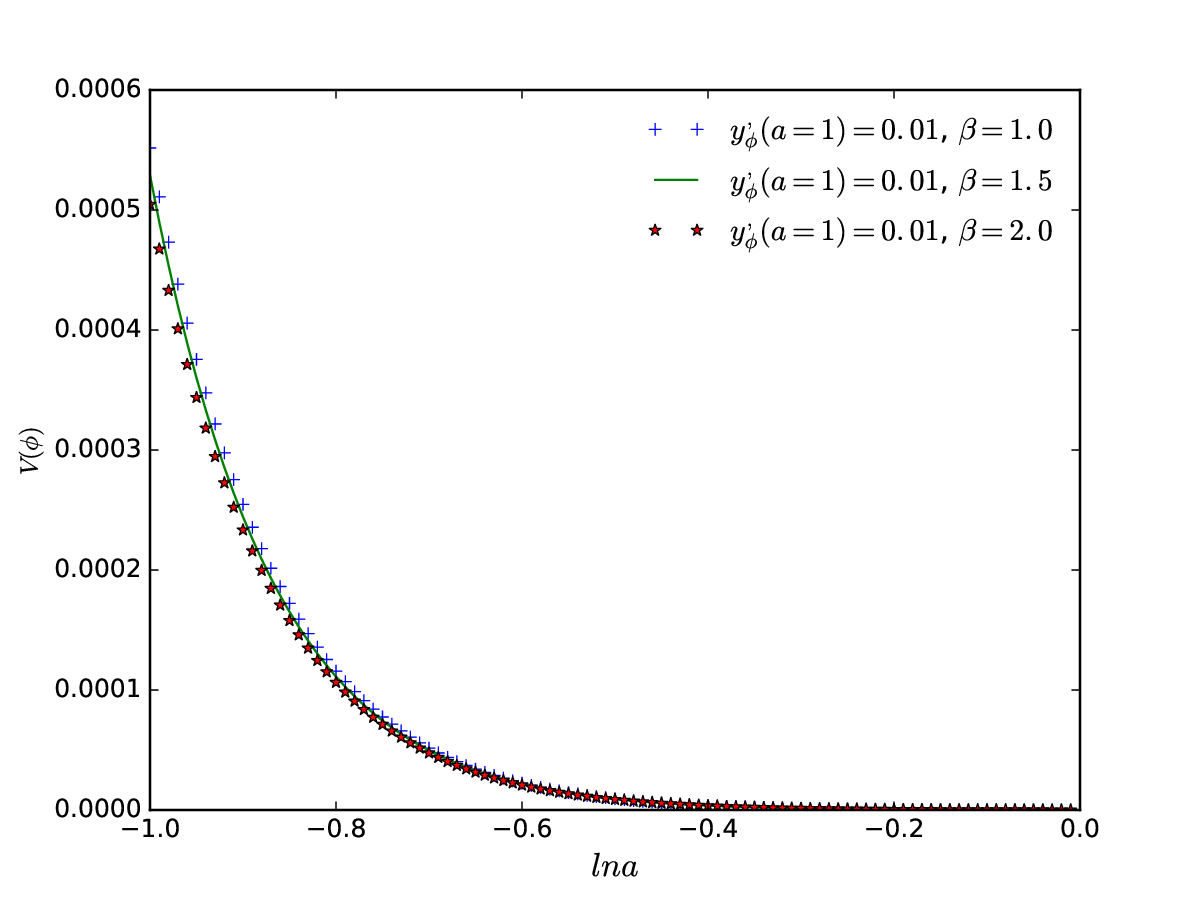}
  \includegraphics[width=7.5cm]{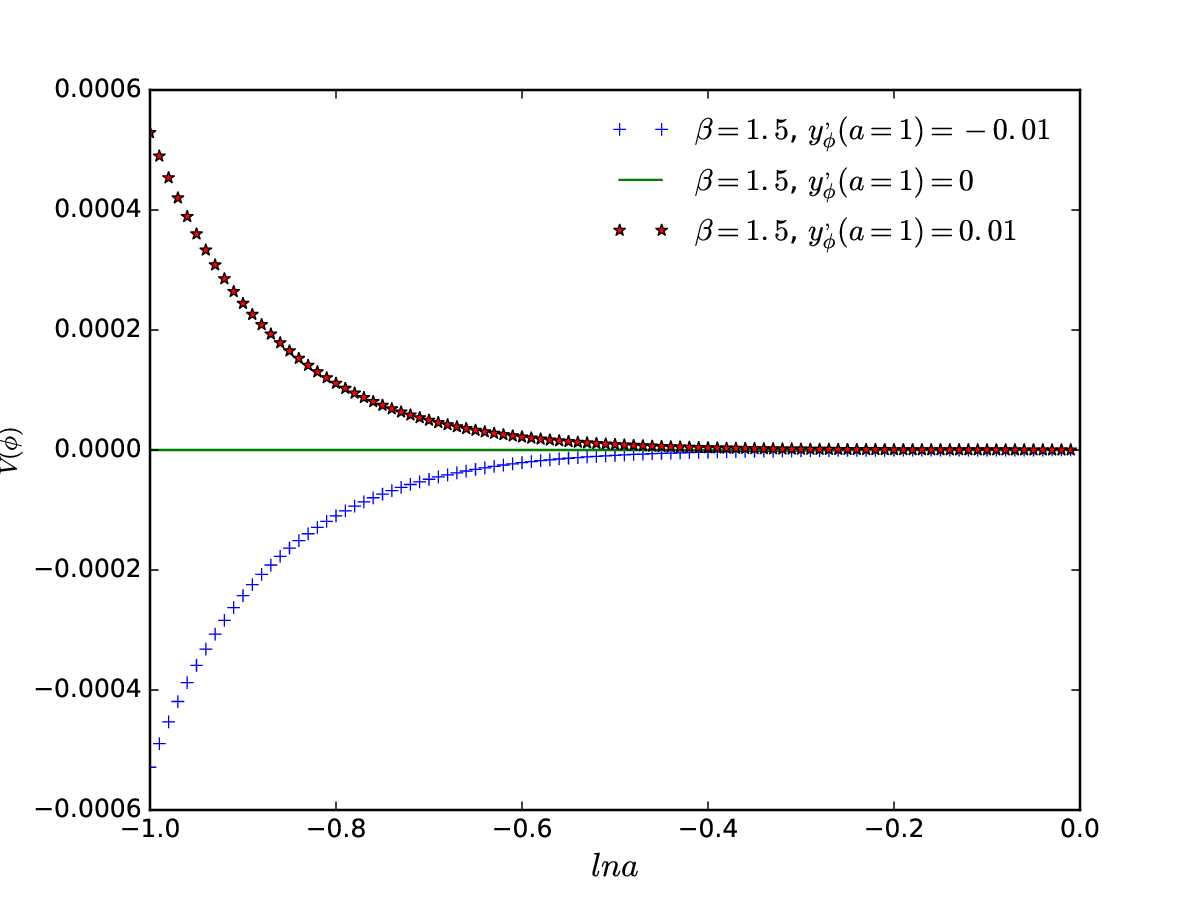}\\
  \caption{The evolutions of effective potential for Brans-Dicke scalar field with the different model parameter $\beta$ and the different  initial condition  $y_{\phi}^{'}(a_{0})$. }\label{figure-vphi}
\end{figure}

\section{$\text{Conclusion}$}

The GBD theory is investigated in this paper, which is obtained by generalizing the Ricci scalar $R$  to an arbitrary function $f(R)$  in the original Brans-Dicke action.  This theory can be reduced to the original  BD theory and the $f(R)$ modified gravity under certain conditions. We give the gravitational field equation and the BD scalar-field  equation in the GBD theory. Using the FLRW metric and the field equations, we can obtain the  cosmological equations in this theory.  The evolutional  equations of universe and BD field are numerically solved by taking a concrete form of $f(R)$ function.  It is shown that the modification to $H(a)$ from  the dynamical BD scalar field is notable, and the GBD model can pass through the test of the observation, such as the observational Hubble data.

 The trajectories of the effective state parameter for the  geometrical dark energy is studied in the GBD universe, which indicates that the evolutions of $w^{eff}_{g}(a)$  with $y_{\phi}^{'}(a_0)\neq 0$   can vary from radiation  ($w^{eff}_{g}\sim 1/3$) to dark energy ($w^{eff}_{g}<0$).  And the modifications to the state parameter from the BD scalar field is remarkable.  In addition, the effective potential of Brans-Dicke field is investigated in the GBD model. One can see that the evolutions  of the BD effective potential  depend  on the initial value of  $y_{\phi}^{'}(a_{0})$, especially they are sensitive to the given symbol of $y_{\phi}^{'}(a_{0})$.

   Ref. \cite{GBD-L} shows an interesting property of the GBD theory, where the GBD theory can naturally solve the problem of $\gamma$ value emerging in $f(R)$ modified gravity (i.e. the inconsistent problem between the observational $\gamma$ (PPN parameter) value and the theoretical $\gamma$ value), without introducing the so-called chameleon mechanism.   In this paper, we also compare our results with other theories. It can be seen that  the GBD theory have some other interesting properties or solve some problems existing in other theories. (1) We can notice that it is required to include both the canonical quintessence field and the non-canonical phantom field in the double scalar-fields quintom model, in order to make the state parameter to cross over the phantom boundary: $w=-1$, while several fundamental problems are associated with the non-canonical phantom field, such as the problem of negative kinetic term and the fine-tuning problem, etc. It can be found that in this paper the effective state parameter of geometrical dark energy in the GBD model can cross over the phantom boundary without bearing the problems relating with the phantom field. (2) It is well known that, the $f(R)$ theory are equivalent to the BDV theory with a specific value of the BD parameter $\omega=0$. However, the specific choice: $\omega=0$ for the BD parameter is quite exceptional, and it is hard to understand the corresponding absence of the kinetic term for the scalar field in the action of the BDV theory, while in the GBD theory the value of $\omega$ is arbitrary and the dynamical effect of the scalar field is non-disappeared. (3) One knows that the geometrical representation may be more appealing to relativists due to its more apparent geometrical nature, whereas the scalar-field representation  seems more appealing to particle physicists. Obviously, the GBD theory tends to investigate the physics from the viewpoint of geometry, while the BDV or the quintom scalar-field model tends to solve physical problems from the viewpoint of matter. Given that the equivalence between the BDV theory and the $f(R)$ theory, some properties of the BD scalar field could be found. So, it is possible that several special characteristics of scalar fields could be revealed through studies of geometrical gravity in the GBD. As shown in this paper, an effective form of the BD potential can be gained by studying the GBD theory. And, it seems that a viable condition for the BD theory could be found, i.e. the BD parameter should be $\omega>0$ for $f>0$, if we assume that the effective form of the BD potential can be approximately written as a popular square function of $\phi$.

\textbf{\ Acknowledgments }
 The research work is supported by   the National Natural Science Foundation of China (11645003,11705079,11575075,11475143).

\end{document}